\documentclass[reprint,prd,amsmath,amssymb,aps,lengthcheck,floatfix,superscriptaddress]{revtex4-2}
\usepackage[utf8]{inputenc}
\usepackage{balance}
\usepackage{graphicx}
\usepackage{dcolumn}
\usepackage{bm}
\usepackage{hyperref} 
\usepackage{amsmath}
\usepackage{slashed}
\usepackage{multirow}
\usepackage{soul}
\usepackage{float}
\usepackage{times}
\usepackage[normalem]{ulem}
\usepackage{color}
\usepackage{cellspace}
\usepackage{lipsum}

\newcommand{\be}{\begin{equation}}
\newcommand{\ee}{\end{equation}}
\newcommand{\bea}{\begin{eqnarray}}
\newcommand{\eea}{\end{eqnarray}}

\newcommand{\comment}[1]{}
\renewcommand\sout{\bgroup \bf \ULdepth=-.5ex \ULset}
\usepackage{xcolor}
\usepackage{orcidlink}
\def\simge{\mathrel{\rlap{\raise 0.511ex
     \hbox{$>$}}{\lower 0.511ex \hbox{$\sim$}}}}
\def\simle{\mathrel{\rlap{\raise 0.511ex
      \hbox{$<$}}{\lower 0.511ex \hbox{$\sim$}}}}

\usepackage{orcidlink}

\begin{document}
\title[Nambu--Jona-Lasinio description of hadronic matter from a Bayesian approach]{Nambu--Jona-Lasinio description of hadronic matter from a Bayesian approach}

\author{K. D. Marquez \orcidlink{0000-0003-0409-9282}} \email{kau@ita.br}
\affiliation{CFisUC, Department of Physics, University of Coimbra, 3004-516, Coimbra, Portugal}
\affiliation{Departamento de Física e Laboratório de Computação Científica Avançada e Modelamento (Lab-CCAM),
Instituto Tecnológico de Aeronáutica, DCTA, 12228-900, São José dos Campos, SP, Brazil}

\author{Tuhin Malik \orcidlink{0000-0003-2633-5821}} \email{tuhin.malik@uc.pt}
\affiliation{CFisUC, Department of Physics, University of Coimbra, 3004-516, Coimbra, Portugal}

\author{Helena Pais \orcidlink{0000-0001-7247-1950}}\email{hpais@uc.pt}
\affiliation{CFisUC, Department of Physics, University of Coimbra, 3004-516, Coimbra, Portugal}

\author {Débora P. Menezes \orcidlink{0000-0003-0730-6689}}  \email{debora.p.m.26@gmail.com}
\affiliation{Departamento de F\'{\i}sica - CFM, Universidade Federal de Santa Catarina, 88.040-900, Florian\'opolis/SC, Brazil}

\author{Constança Providência \orcidlink{0000-0001-6464-8023}}
\email{cp@uc.pt}
\affiliation{CFisUC, Department of Physics, University of Coimbra, 3004-516, Coimbra, Portugal}

\date{\today}

\begin{abstract}
A microscopic nuclear matter formalism with explicit chiral symmetry based on the Nambu Jona-Lasinio model is considered to describe nuclear matter. To reproduce nuclear matter properties adequately at the saturation density, four-point and eight-point interactions are introduced. Within a Bayesian inference approach, the parameters of the model are determined by imposing nuclear matter, both experimental and from {\it ab-initio} calculations, and neutron star observational constraints.  Nuclear matter properties are well reproduced with an effective mass of 0.75 to 0.8 nucleon mass at the saturation density. At 90\% confidence level, the radius of a {$1.4 ~\rm M_\odot$} star varies between 11.48~km and 13.20~km,  masses as large as $\sim 2.2 ~\rm M_\odot$ are predicted and the radius of a 2 M$_\odot$ star is above 10.5~km. High-density perturbative QCD ({pQCD}) results exclude equations of state that predict larger maximum masses and radii. The speed of sound increases monotonically with density and reaches values as large as $\sqrt{0.7}c$--$\sqrt{0.8}c$ in the center of massive stars. Several properties such as the polytropic index or the renormalized trace anomaly, that have been proposed to identify the deconfined phase transition, are analyzed. Interestingly, the radius of the obtained posterior that also meets pQCD constraints aligns closely with the mass-radius measurement of the recent PSR J0437-4715, which contrasts with other relativistic mean field model results.
\end{abstract}

\maketitle

\section{Introduction}\label{sec1}
To the current understanding of the physical processes given by the Standard Model, the fundamental theory of the strong interaction is considered to be Quantum Chromodynamics (QCD). It shows non-perturbative behaviors at low energies. Among these, the most relevant for the discussion we intend to propose in this work is the nontrivial vacuum structure which arises from scalar quark-antiquark pairs leading to a finite expectation value of quark condensates $\langle\bar qq\rangle$. 
Since the initial QCD Lagrangian exhibits a $SU(2)_L\times SU(2)_R$ group symmetry if quarks were taken as massless particles in the tree level, this non-vanishing $\langle\bar qq\rangle$ property implies that the quark masses can be considered to be dynamically generated by the spontaneous chiral symmetry breaking. As quarks do have a finite current mass, the chiral symmetry of QCD is broken explicitly by these terms. However, the chiral symmetry can still be seen as an approximate symmetry of the theory because current quark masses are small when compared to the interaction scale. It renders, through spontaneous breaking, light but not massless {pseudo-Goldstone bosons} that can be identified as the three pions (when only $u$ and $d$ quarks are considered) \cite{Peskin}.
 
The non-perturbative nature of the strong interaction also makes it necessary to resort to other methods when describing hadronic matter, be either nucleon matter near the nuclear saturation density or stellar matter at higher densities, as the one expected in the center of compact objects. The description of nuclei, nuclear and stellar matter through effective models can be performed constructing field theories that, ideally, would include QCD symmetries in the Lagrangian density of the system \cite{waleckabook,buballa}.

Walecka-type hadrodynamics models \cite{walecka,nlwmbb} are {amongst} the most popular choices, especially in the nuclear astrophysics literature. In an extensive review, Ref. \cite{nlwm} analyzed 263 representative models of this class and compared their predictions with well-established experimental quantities for nuclear matter and stellar properties. Theoretical results were found to agree with empirical limits for nuclear matter quantities in only 35 of these models, i.e. those that (i) have coupling constants dependent on the baryon density and/or (ii) include cross-interaction terms of the mesonic fields. After considering the more restrictive astrophysical constraints, Ref. \cite{nlwmstellar} reduced these 35 valid models to only 2 options, both of which fall into category (i). Nevertheless, in the decade between that study and today, this type of model has continued to be widely accepted in the community, with many newer parameterizations being proposed that already take into account the more restrictive experimental constraints -- especially after the LIGO/Virgo Collaboration gravitational wave detection of neutron star mergers \cite{LIGOScientific:2018cki,Abbott2020} and the Neutron Star Interior Composition Explorer (NICER) measurements of neutron star properties through soft x-ray timing \cite{Riley2019,Miller2019,Riley2021,Miller2021} began to provide many more quantitative results on neutron stars properties. However, as the chiral symmetry is a fundamental characteristic of the strong force 
incorporating chiral symmetry breaking/restoration in effective nuclear models ensures theoretical consistency with the underlying principles of QCD, and Walecka-type models do not consider it explicitly.

The first theoretical approach to the chiral symmetry breaking in a fermionic quantum field theory was given by the seminal 1961 papers of Y. Nambu and G. Jona-Lasinio  \cite{Nambu:1961tp}, anticipating the proposal of quarks and the QCD formulation itself. The original {Nambu--Jona-Lasinio} (NJL) model aimed to describe interacting nucleons, motivated by the fact that this interaction holds a partially conserved {axial vector current}, associated with the approximate chiral symmetry. Since it imposes the nucleon mass to be small in the Lagrangian level, the spontaneous symmetry breaking provides a mechanism that generates dynamically the large nucleon mass. Even though the original NJL model was proposed considering nucleonic degrees of freedom,  the lack of confinement soon became a problem within the QCD theoretical framework, which made it to be reinterpreted as a schematic quark model \cite{Kleinert1978}. The fact that the NJL Lagrangian does not confine the fermion fields also implies that it is unable to describe the saturation properties of the nuclear matter. However, this issue was fixed by the inclusion of an extra 8-point interaction channel representing a crossed scalar-vector interaction \cite{Koch1987}. 
The same logic can be applied to many other channels of the interaction, aiming to improve the description of other important bulk properties, which give origin to the so-called {extended Nambu--Jona-Lasinio} (eNJL) models family \cite{enjl,Wei:2015aep}. 
{ This idea of an 8-point interaction has also been applied to the description of quark matter, see Refs. \cite{benic2014heavy,benic2015new,kaltenborn2017quark,Ferreira:2020kvu}}. 

Bayesian methods have gained popularity for mapping the parameter space of dense matter effective models \cite{Imam:2021dbe,Malik:2022ilb,Coughlin:2019kqf,Wesolowski:2015fqa,Furnstahl:2015rha,Ashton:2018jfp,Landry:2020vaw,Char:2023fue,Imam:2024gfh,Huang:2023grj,Scurto:2024ekq,Zhu:2022ibs,deMouraSoares:2023cqq,Traversi:2020aaa}. The Bayesian inference offers a robust framework for analyzing the parameter space of neutron star equations of state (EoS), enabling a thorough understanding of the uncertainties and probabilistic constraints linked to the physical properties of these dense objects. By integrating prior knowledge with observational data, Bayesian techniques enable the construction of posterior probability distributions for EoS parameters like pressure, density, and composition at various depths within a neutron star. This method not only helps in limiting the range of feasible EoS models but also provides quantification of the confidence levels in these constraints. Additionally, nonparametric Bayesian approaches, including those utilizing feed-forward neural networks (FFNN), offer versatile tools for reconstructing the EoS from multimessenger observations. These techniques are not dependent on predefined functional forms, enabling them to accommodate a vast array of theoretical EoS models and observational data \cite{Han:2021kjx}. 
Moreover, several EoS metamodels, constrained by {\it ab-initio} theoretical calculations for both low and high densities, have been proposed: nucleon-nucleon chiral potentials for low-density neutron and nuclear matter \cite{Hebeler:2013nza,Drischler:2015eba} and perturbative Quantum Chromodynamics for extremely high-density regimes \cite{Kurkela:2009gj}. To reconcile all possible EoS compatible with these two constraints, the EoS at the extreme densities are connected using either piecewise polytropic interpolation, speed-of-sound interpolation, or spectral interpolation, with causality constraints imposed as needed \cite{Lindblom:2012zi,Kurkela:2014vha,Most:2018hfd,Oter:2019rqp,Annala:2021gom}. Recently, a nonparametric inference of the neutron star EoS has also been proposed based on Gaussian processes \cite{Essick:2019ldf} or using machine learning techniques \cite{Han:2021kjx}. However, such EoS models have significant limitations as they do not assume any specific composition of matter in the intermediate density regime. 

In this work, a generalized version of these eNJL models is presented for the case of nucleonic (protons and neutrons) dense matter, to be applied in the description of symmetric and pure neutron matters around the nuclear saturation density and, also, of the stellar matter in compact objects such as neutron stars. The study will be performed within a Bayesian inference approach to determine the model parameters given a set of experimental and observational nuclear matter and neutron star properties, following closely the procedure applied in \cite{Malik:2022zol,Malik:2023mnx}. In particular, we will also consider the perturbative QCD constraints following the arguments discussed in Ref. \cite{Komoltsev:2021jzg}, that delineates the equation of state behavior necessary to be consistent with perturbative QCD results at high densities.  Bayesian inference methodologies have been intensively applied to constrain the hadron EoS given a phenomenological microscopic model in the last years, see for instance \cite{Traversi:2020aaa,Malik:2022zol,Beznogov:2022rri,Malik:2023mnx,Char:2023fue,Scurto:2024ekq,Imam:2024gfh}, and in this study we aim to apply this robust methodology for the first time to a hadronic model within the extended Nambu--Jona-Lasinio model framework.

The paper is organized as follows: in Sec. \ref{sec11} the eNJL formalism is reviewed, the nuclear matter properties introduced and the Tolmann-Oppenheimer-Volkof equations to determine the neutron star spherical equilibrium configurations summarized; in Sec. \ref{sec3} the Bayesian inference methodology applied in the present study is introduced; in Sec. \ref{results}, we present the results of our Bayesian calculation, in particular, the nuclear matter properties of the EoS data set and the respective neutron star properties; we finish by summarizing the main results and drawing some conclusions in Sec. \ref{conclusion}.

\section{Hadronic extended Nambu--Jona-Lasinio formalism}\label{sec11}
In this study, we use the eNJL model proposed initially by Ref.~\cite{enjl}. This rather broadened eNJL formalism can describe hadronic matter constituted by protons and neutrons, and its Lagrangian density is given by
\begin{equation}
 \mathcal{L}_{\rm eNJL}=\bar{\psi}\left(i\slashed{\partial}-\hat m\right)\psi+\mathcal{L}_I. \label{eq:enjlgen}
\end{equation}
In (\ref{eq:enjlgen}),  $\psi=\left[\psi_p\; \; \; \psi_n\right]^T$ represents the nucleon fields of masses $\hat m=\text{diag}(m,m)$, assuming isospin symmetry in the Lagrangian level {($m_p=m_n=m$)}. The interaction part $\mathcal{L}_I$ is given by several four-point and  eight-point interactions compatible with the SU(2) flavor symmetry, written as
\begin{align}
\mathcal{L}_I&= 
G_s[(\bar{\psi}\psi)^2 + (\bar{\psi}i\gamma_5\vec{\tau}\psi)^2]  - G_v(\bar{\psi}\gamma^\mu\psi)^2 \nonumber \\
&- G_{sv}[(\bar{\psi}\psi)^2 + (\bar{\psi}i\gamma_5\vec{\tau}\psi)^2](\bar{\psi}\gamma^\mu\psi)^2   
 \nonumber\\&-G_\rho[(\bar{\psi}\gamma^\mu\vec{\tau}\psi)^2 + (\bar{\psi}\gamma_5\gamma^\mu\vec{\tau}\psi)^2]
  \nonumber\\	& - G_{s\rho} [(\bar{\psi}\psi)^2 + (\bar{\psi}i\gamma_5\vec{\tau}\psi)^2] [(\bar{\psi}\gamma^\mu\vec{\tau}\psi)^2 + (\bar{\psi}\gamma_5\gamma^\mu\vec{\tau}\psi)^2] \nonumber\\&-G_{v\rho}(\bar{\psi}\gamma^\mu\psi)^2[(\bar{\psi}\gamma^\mu\vec{\tau}\psi)^2 + (\bar{\psi}\gamma_5\gamma^\mu\vec{\tau}\psi)^2]
  , \label{eq:henjl}
\end{align}
with $G_a, a=s,v,\rho$ standing for the coupling constants for the different channels, $G_{ab}, a\neq b$, representing the crossed interactions between channels $a$ and $b$, and where $\vec{\tau}$ are the Pauli isospin matrix. {As already mentioned, for} nuclear matter, the simpler versions of the NJL model with four-point interactions do not lead to binding. In Ref. \cite{Koch1987}, the term $G_{sv}$ was firstly introduced  to overcome this problem.
The $G_v$ term simulates a chiral-invariant short-range repulsion between the nucleons, the $G_\rho$ term  allows for the description of isospin asymmetric matter, the $G_{sv}$ term accounts for a density dependence of the scalar coupling, and the $G_{s\rho}$ and $G_{v\rho}$ terms make the density dependence of the symmetry energy softer.  
As a consequence of the dimensionality of the coupling constants, NJL-type models are not renormalizable, thus a regularization scheme must be employed to the divergent integrals. Here, it is done by applying a sharp cutoff in the three-momentum ultraviolet region of the above integration limits, denoted as $\Lambda$. This parameter is calculated such that the nucleon mass in the vacuum is 939 MeV.

The vacuum expectation $\langle 0|\bar{\psi}\psi|0\rangle$ is non-zero, thus responsible for the breaking the chiral symmetry.  The  equation for the spinor field can be obtained through the so-called {Hubbard-Stratonovich transformation}, i.e., the bosonization of the model by auxiliary fields given by the non-vanishing condensates \cite{buballa}. 
Neglecting quadratic terms in the fluctuations, the Lagrangian is linearized by taking 
\begin{align}
(\bar\psi\psi)^2={}&2n_s(\bar\psi\psi)-n_s^2,\\
(\bar\psi\gamma^\mu\psi)^2={}&2n_B(\psi^\dagger\psi)-n_B^2, 
\end{align}
where the scalar and vector condensates, also called densities, are given as $n_s=\left\langle\bar\psi\psi\right\rangle={n_s}_p+{n_s}_n$ and $n_B=\left\langle\psi^\dagger\psi\right\rangle=n_p+n_n$.
 
In particular, terms in channels without condensate or the space components in the vector vertex drop out. From the variational principle, one obtains the {Dirac-like equation}
\begin{equation}
\left(i\slashed{\partial}-M^*\right)\psi=0,
\end{equation}
where this shift in the mass by the non-zero condensates can be understood as a non-perturbative correction to the self-energy, i.e., to consider the fermion propagator {dressed} by the one-loop correction. 
The particles governed by the Lagrangian (\ref{eq:henjl}) behave as non-interacting particles with mass $M^*\ne m$ and an effective chemical potential $\tilde{\mu}\ne\mu$.

The thermodynamic potential per volume of the hadronic eNJL model is obtained directly from the Lagrangian density (\ref{eq:henjl}), and reads 
\begin{align}
{\Omega}={}&\sum_{i = p,n} (\epsilon_{i}-\mu_i n_i )+m n_s-G_sn_s^2+G_vn_B ^2\nonumber
\\ &+G_{sv}n_s ^2n_B ^2+G_\rho n_3^2 +G_{s\rho} n_s ^2n_3^2+G_{v\rho}n_B ^2n_3^2  ,  \label{eq:omegahenjl}    
\end{align}
where $m$ is the current mass that is introduced in the models to make the restoration of the chiral symmetry softer. The nucleon effective mass, $M^*$, comes from the condition $\partial \Omega/\partial M^*=0$, and is equal to
\begin{equation}
        M^* =m - 2G_sn_s + 2G_{sv}n_sn_B^2 + 2 G_{s\rho}n_sn_3^2  . \label{eq:gap1}
\end{equation}
The chemical potential, $\mu_i$, is determined by imposing $\partial \Omega/\partial k_{F{_i}}=0$ and is given by
\begin{align}
	\mu_i ={}& E_{F_i} +2G_vn_B + 2G_{sv}n_Bn_s^2   \pm 2G_\rho n_3 \nonumber\\
 &\pm 2 G_{s\rho}n_3n_s^2 + 2G_{v\rho}n_Bn_3(n_3\pm n_B )  ,
\end{align}
with the upper (lower) signs taken for $i=p$ ($n$), and $n_3=n_p-n_n$ is the isovector density.
 
The $T=0$ limit is a reasonable approximation in situations where the thermal energy is smaller than the average kinetic energy of the particles of the system, i.e., when $k_BT\ll E_p$, as is the case inside compact stars with temperatures of the order of $10^{10}$ K or below.
In this regime, the nucleon kinetic energy density $\epsilon_i$ is given by 
\begin{align}
  \epsilon_i=F_1(M^*,k_{F_i})-F_1(M^*,\Lambda)  ,  
\end{align}
where
\begin{equation}
F_1(M^*,x)=\int_{0}^{x } \frac{dk}{ \pi ^2}k^2\sqrt{k^2+M^{*2}}.
\end{equation}
The Fermi energy, which at $T=0$ corresponds to the most energetic occupied level, is defined as
\begin{equation}
{E_F}_i=\tilde{\mu}_i=\sqrt{{k_F}_i^2+M^{*2}},\label{eq:ur}
\end{equation}
where ${k_F}_i$ is the Fermi momentum, for $i=p,n$. It allows to define the nucleon distribution functions as $f_i=\theta({k_F}_i-k)$, hence, follows the number density of a degenerate Fermi gas being 
\begin{equation}
{n_B}_i=2\int_{0}^{k_F }  \frac{d^3k}{\left ( 2\pi  \right )^3}\theta({k_F}_i-k)=\frac{1}{3\pi^2}{k_F}_i^3, \label{eq:rhopf}
\end{equation}
where ${k_F}_i=\sqrt{\tilde{\mu}_i^2-M^{*2}}$, always demanding ${k_F}_i\leq\Lambda$. Also, the scalar density can be written as
\begin{equation}
    {n_s}_i=M^*[F_0(M^*,{k_F}_i)-F_0(M^*,\Lambda )],  \label{eq:rhospf}
\end{equation}
where the function {$F_0(M^*,x)$ is defined as}
\begin{equation}
F_0(M^*,x)=\int_{0}^{x } \frac{dk}{ \pi ^2}\frac{k^2}{\sqrt{k^2+M^{*2}}}.
\end{equation}

In general, the gap equation \eqref{eq:gap1} has more than one solution, e.g., in the chiral limit the trivial configuration $M^*=0$ is always a valid solution, but non-trivial solutions $M^*=\pm M_0\neq 0$ can occur as well. It is possible to show that the vacuum energy is always minimized by the solution with the largest value of $M^*$ \cite{vogl}. In this situation, where there is a medium, the occupation numbers are non-zero and reduce the value of the constituent mass for low temperatures and densities. As the temperature or the density increases, the particle distribution factor tends to zero, and the constituent mass $M^*$ approaches the value of the current mass $m$. This mechanism is called chiral symmetry restoration, as the ground state solution goes back to the light particle case.

It is straightforward to write the equations of state (EoS) for the hadronic matter at $T=0$ from the thermodynamic potential, through the fundamental relations $P=-\Omega + \epsilon_0$, with $\epsilon_0$ being the energy density in the vacuum, $P$ the pressure, and
\begin{equation}
    \varepsilon=\sum_{i = p,n}{\mu }_i n_i-P  ,
\end{equation}
the energy density. The application of the nucleon matter EoS obtained above to the description of bulk nuclear matter quantities from the expressions of $P$ and $\varepsilon$ in the zero-temperature limit is discussed in the following.

It is convenient to define the {asymmetry coefficient} in terms of the number density of the individual particle species 
\begin{equation} 
\alpha =\frac{{n_B}_n-{n_B}_p}{{n_B}},
\end{equation}
such that
${k_F}_p=k_F\left ( 1-\alpha  \right )^{1/3}$
and
${k_F}_n=k_F\left ( 1+\alpha  \right )^{1/3},$
The symmetric matter case is reached simply by taking ${\alpha=0}$.

The {saturation density} $n_0$ is defined as the density of symmetric nuclear matter where the {binding energy} $\mathcal{E}(n_B,\alpha)=\varepsilon/n_B-M_0$ reaches its minimum, with $M_0=939 ~\rm MeV$ the nucleon mass in vacuum, i.e., when 
\begin{equation} 
\left.\frac{\partial \mathcal{E}}{\partial n_B}\right|_{n_B=n_0}\!\!\!=0,
\end{equation}
with $\alpha =0$.  
Several of the following bulk quantities are known experimentally or constrained theoretically at the saturation density of nuclear matter, or in some band region around $n_0$ (see, e.g., \cite{nlwm,syme2} and references within). The index in $\mathcal{O}_0$ indicates  the quantity $\mathcal{O}$ taken at $n_B=n_0$.
From the EoS, i.e., $P$ and $\varepsilon$, it is possible to calculate the {incompressibility modulus} \cite{kzero}
\begin{equation} 
K_0=9\left.\frac{\partial P}{\partial n_B}\right|_{\alpha=0}\! \! \! ,
\end{equation}
and the skewness and kurtosis coefficients, respectively given by
\begin{equation}
Q_0=27n_{0}^3\left.\frac{\partial^3(\varepsilon/n_B)}{\partial n_B^3}\right|_{\alpha=0}\! \! \! ,
\end{equation} 
and
\begin{equation}
Z_0=81n_{0}^4\left.\frac{\partial^4(\varepsilon/n_B)}{\partial n_B^4}\right|_{\alpha=0}\! \! \! . 
\end{equation}
Another set of bulk nuclear matter parameters follows from the symmetry energy, which is an important quantity to model nuclear matter and finite nuclei by probing the isospin part of nuclear interactions, given by
\begin{equation}
\mathcal{S}=\frac{1}{2}\left. \frac{\partial ^2(\varepsilon/n_B)}{\partial \alpha ^2} \right |_{\alpha =0}. \label{eq:esym}
\end{equation}
 One can expand the symmetry energy $\mathcal{S}$ around the saturation density $n_0$ as
\begin{align} 
\mathcal {S}(n_B )={}&\mathcal {S}_0+L_{\rm sym,0}\eta +\frac{1}{2}K_{\rm sym,0}\eta ^2+\frac{1}{6}Q_{\rm sym,0}\eta ^3\nonumber\\ &+\frac{1}{24}Z_{\rm sym,0}\eta^4+\mathcal {O}(\eta ^5),
\end{align}
where $\eta(n_B )=(n_B-n_0 )/3$ and the coefficients of the expansion are, respectively, the {slope of the symmetry energy},
\begin{equation} 
L_{\rm sym,0}=3n_{0}\frac{\partial \mathcal{S}}{\partial n_B}
\end{equation}
the {curvature of the symmetry energy},
\begin{equation} 
K_{\rm sym,0}=9n_{0}^2\frac{\partial^2 \mathcal{S}}{\partial n_B^2},
\end{equation}
the {skewness of the symmetry energy},
\begin{equation} 
Q_{\rm sym,0}=27n_{0}^3\frac{\partial^3\mathcal{S}}{\partial n_B^3},
\end{equation}
and the {kurtosis of the symmetry energy},
\begin{equation} 
Z_{\rm sym,0}=81n_{0}^4\frac{\partial^4\mathcal{S}}{\partial n_B^4},
\end{equation}
with all derivatives taken at $n_B=n_0$ \cite{esymreview}.

The presented phenomenological formalism relies on experimental or observational data for calibration. It is then essential to fit the model to nuclear matter properties at the saturation density and to pure neutron matter quantities because neutron star matter is expected to be highly asymmetric, for which pure neutron matter serves as an approximation. Neutron matter can be derived from ab initio theoretical calculations for low-density neutron and nuclear matter, such as the N$^3$LO calculation in chiral effective field theory ($\chi$EFT) \cite{Hebeler:2013nza}. Complementarily, for the extremely high-density regimes, perturbative Quantum Chromodynamics (pQCD) has been considered as a high-density constraint to the phenomenological models, because the effective (non-perturbative) models should conform to pQCD in this limit \cite{Kurkela:2009gj}.

The procedure for moving from microphysics, in the form of effective relativistic models, to macrophysics, in the form of observational variables of compact stellar objects, will be presented in the following.
To fulfill the charge neutrality and chemical equilibrium constraints of the stellar matter, a non-interacting electron gas is included in the description. Then, the particle fractions are determined from charge neutrality, $n_p=n_e+n_\mu$, where $n_e$ ($n_\mu)$ is the number densities of electrons (muons), and from the $\beta$-equilibrium condition, expressed in terms of the neutron and electron chemical potentials, 
$\mu_p=\mu_n-\mu_e,$
with $\mu_\mu=\mu_e$. 

The crust was constructed using the Baym-Pethcik-Sutherland EoS \cite{Baym:1971pw} to describe the outer crust and a polytropic function to describe the inner crust. This function connects the outer crust EoS to the core hadronic eNJL stellar EoS at $n_B=0.04$ fm$^{-3}$. The choice of the matching density follows the steps proposed in \cite{Malik:2023mnx} and tested in \cite{Malik:2024nva}, where it was shown that the uncertainty introduced in the determination of the radii of low-mass stars is not greater than $0.1$ km. {After the Bayesian inference is performed, a unified inner crust-core EoS were built according to the approach discussed in \cite{enjl} for some sample posterior parameterizations.}

Moving from micro to macrophysics requires submitting the EoS that describes dense matter to conditions of mechanical (or hydrostatic) equilibrium since compact stars are understood to be sufficiently stable objects in their internal structure. Compact stars are bodies whose gravitational field is extremely intense, so the equilibrium relationship must be established within the framework of general relativity. The equation for relativistic hydrostatic equilibrium (\ref{eq:tov}) is called the Tolman-Oppenheimer-Volkoff (TOV) equation, and reads
\begin{equation}
 \frac{d P}{d r}=-\frac{\left [ \varepsilon \! \left ( r \right )+P\! \left ( r \right ) \right ]\left [ m\! \left ( r \right )+4\pi r^{3} P\! \left ( r \right ) \right ]}{r\left [ r-2m\! \left ( r \right ) \right ]}, \label{eq:tov}
\end{equation}
where the gravitational mass is
\begin{equation}
 m \left ( r \right )=\int_{0}^{r}d r{}'4\pi{{r}'^{2}}\varepsilon\! \left ( r{}' \right ), \label{eq:tovm}
\end{equation}
and $\varepsilon \! \left ( r \right )$ and $P\! \left ( r \right )$ are the energy density and pressure in the spherical shell of radius $r$.
This object is perceived by a distant observer as having a radius $r=R$, defined from the boundary condition $P(R)=0$, and a gravitational mass $M=m(R)$, given by (\ref{eq:tovm}). Other boundary conditions are also important, such as $m(0)=0$ and the definitions of central pressure and central energy density,
$ P\! \left ( 0 \right )=P_c $ and $  \varepsilon\! \left ( 0 \right )=\varepsilon_c$.

Following the first NS binary merger measured by the LIGO/VIRGO collaboration in 2017 \cite{LIGOScientific:2018cki}, study of relativistic tidal effects in compact stars has been the focus of intense research \cite{Malik:2018zcf,Chatziioannou:2020pqz,Baiotti:2019sew,Carson:2018xri}. The dimensionless tidal parameter $\Lambda$, which measures how easily an object is deformed by an external gravitational field, is of special interest, as this parameter is measured by gravitational wave observations and can be also confronted against theoretical model predictions. We refer to \cite{Hinderer:2007mb} to a review on the calculation of the tidal deformability parameter.

\section{Inference Framework \label{sec3}}
The Bayesian method for estimating parameters is a reliable statistical approach used to evaluate model parameters from a specific dataset \cite{Malik:2022zol,Imam:2021dbe,Wesolowski:2015fqa,Furnstahl:2015rha,Ashton:2018jfp,Huang:2023grj,Char:2023fue,Imam:2024gfh}. By leveraging Bayes' theorem, one can integrate prior beliefs about the parameters with new data, yielding posterior distributions that encapsulate updated knowledge about the parameters. The posterior distributions of the model parameters $\boldsymbol{\theta}$ in Bayes’ theorem can be written as
\begin{equation}
 P\left (\boldsymbol{ \theta} | D \right )=\frac{\mathcal{L}\left ( D|\boldsymbol{ \theta}  \right )P\left ( \boldsymbol{ \theta} \right )}{\mathcal{Z}}, \label{eq:bayestheorem}
\end{equation}
with $\boldsymbol{ \theta}$ and $D$ denoting the set of model parameters and the fit data, respectively. In Eq. \eqref{eq:bayestheorem}, $P\left ( \boldsymbol{ \theta} \right )$ represents the prior distribution for the model parameters, while $\mathcal{Z}$ denotes the evidence. Choosing a prior distribution depends on the initial understanding of model parameters. A uniform prior, often a standard baseline, is frequently used. The joint posterior distribution of the parameters, denoted as $P\left (\boldsymbol{ \theta} | D \right )$, is determined by the likelihood function 
$\mathcal{L}\left ( D|\boldsymbol{ \theta}  \right )$. The posterior distribution of a specific parameter can be derived by the marginalization process, i.e., by integrating 
$ P\left (\boldsymbol{ \theta} | D \right )$ across the remaining parameters. 

The likelihood functions for various quantities analyzed in this study are listed {next}. 
For the symmetric nuclear matter (SNM) properties, we use a Gaussian likelihood, given by
    \begin{equation}
    \mathcal{L}^{\rm SNM}(D |\boldsymbol{\theta})=\prod_j \frac{1}{\sqrt{2 \pi \sigma_j^2}} e^{-\frac{1}{2}\left(\frac{d_j-m_j(\boldsymbol{\theta})}{\sigma_j}\right)^2} , \label{eq:gausslikelihood}
    \end{equation}
 where the index $j$ runs over all the datapoints (see table \ref{tab:constraints}), $d_j$ are the constraining data values, $m_j(\boldsymbol{\theta})$ are the model values corresponding to the set of model parameters $\boldsymbol{ \theta}$, and the $\sigma_j$ are the adopted uncertainties of the data.
The $\chi$EFT constraints to the pure neutron matter (PNM) were enforced using a super-Gaussian box function probability with a minor tail, expressed as,
\begin{align}
    \mathcal{L}^{\rm PNM}(D |\boldsymbol{\theta}) =\prod_j \frac{1}{2 \sigma_j^2}\frac{1} {\exp{\left(\frac{|d_j-m_j(\boldsymbol{\theta})|-\sigma_j}{0.015}\right)}+1} , \label{eq:boxlikelihood}
\end{align}
where $d_j$ is the median value and $\sigma_j$ represents two times the uncertainty of the $j^{\rm th}$ data point from the $\chi$EFT constraints \cite{Hebeler:2013nza}. 
To determine the likelihood associated to the neutron star (NS) maximum mass, we employ a Fermi-Dirac likelihood, ensuring the NS maximum mass exceeds 2 M$_\odot$, and include a light tail to facilitate better sampling convergence, 
    \begin{equation}
        \mathcal{L}^{\rm NS}(D |\boldsymbol{\theta}) = \frac{1}{\exp \left({\frac{m(\boldsymbol{\theta})-2.0}{-0.01}} \right) + 1}.
    \end{equation}
Finally, the pQCD constraints are implemented by the likelihood function
\begin{align}
    {\mathcal L}({d_{\rm pQCD} | }\theta) =P(d_{\rm pQCD}|\theta)= \mathcal{L}^{\rm pQCD} ,
\end{align}
where $d_{pQCD}$ is a constant probability distribution in the energy density and pressure plane at 7$n_0$ (at the renormalizable scale $X=2$), with $n_0=0.16$ fm$^{-3}$, having $P(d_{\rm pQCD}|\theta) = 1$ if it is within $d_{\rm pQCD}$ and zero otherwise. 
So the total likelihood in the inference analysis is given by
\begin{equation}
    \mathcal{L}^{\rm total}=\mathcal{L}^{\rm SNM} \times \mathcal{L}^{\rm PNM} \times \mathcal{L}^{\rm NS} \times \mathcal{L}^{\rm pQCD}.
\end{equation}

It should be noted that other constraints were not applied during the sampling of the inference analysis, such as imposing $\partial \mathcal{E}/\partial n_B>0$ for $n_B>n_0$ and {${\cal S}(n_B)>0$ in SNM, and $\partial P/\partial n_B>0$ in PNM. These conditions were instead used as filters in the resulting posterior and eliminated less than 3\% of the samples from the final posterior. 
For the present study, we used a nested sampling approach using the \texttt{PyMultiNest} \cite{Buchner:2014nha} algorithm with 1500 live points. This resulted in approximately 4000 samples in the final posterior, demonstrating good posterior health.

\section{Results and Discussions}\label{results}
In this section, we discuss the properties of the hadronic dense matter EoS and the corresponding NS {properties} that result from the eNJL model presented above, with its coupling constants constrained by the Bayesian inference process.

\begin{table}[!t]
\centering
\caption{Constraints imposed in the dataset within the Bayesian inference framework: symmetric nuclear matter (SNM) binding energy per nucleon $\mathcal{E}_0$, incompressibility modulus $K_0$, and symmetry energy  $\mathcal{S}_0$, at the nuclear saturation density $n_0$;
pure neutron matter (PNM) pressures determined at baryon densities of 0.08, 0.12 and 0.16~fm$^{-3}$; neutron star (NS) maximum mass $M_{\rm max}$; and perturbative Quantum Chromodynamics (pQCD) derived constraints of the EoS in density relevant for NS ($n_B=7n_0$) for pQCD renormalization scale $X=2$.}
\label{tab:constraints}
  \setlength{\tabcolsep}{0.2pt}
      \renewcommand{\arraystretch}{1.1}
\begin{tabular}{cccc}
\hline
\hline
\multicolumn{2}{c}{Quantity}                      & Constraint                                    & Ref.                                                     \\ \hline
\multirow{6}{*}{SNM} & $n_0$                      & $0.153\pm0.005$ MeV                           & \cite{Typel1999}                        \\
                     & $\mathcal{E}_0$            & $-16.1\pm0.2$ MeV                             & \cite{Dutra2014}                        \\
                     & $K_0$                      & $230\pm40$ MeV                                & \cite{Shlomo:2006ole,fsugold}           \\
                     & $\mathcal{S}_0$            & $32.5\pm1.8$ MeV                              & \cite{Essick:2021ezp}  \\
                     & $\partial \mathcal{E}/\partial n_B$ & $>0$ for $n_B>n_0$ & \\
                     &{${\cal S}(n_B)$} &$>0$ &\\
\multirow{4}{*}{PNM} & $P(n_B=0.08$ fm$^{-3})$     & $0.521\pm0.091$ MeV fm$^{-3}$                 & \multirow{3}{*}{\cite{Hebeler:2013nza}} \\
                     & $P(n_B=0.12$ fm$^{-3})$     & $1.262\pm0.295$ MeV fm$^{-3}$                 &                                                          \\
                     & $P(n_B=0.16$ fm$^{-3})$     & $2.513\pm0.675$ MeV fm$^{-3}$                 &                                                          \\
                     &{$\partial P/\partial n_B$}&$>0$&                \\
NS                   & $M_{\rm max}$              & $>2 ~\rm M_\odot$                                   & \cite{Demorest2010,Antoniadis2013}  \\
pQCD                 & EoS at $n_B=7n_0$ & see Fig. \ref{fig:eos_pQCD} & \cite{Komoltsev:2021jzg}     \\ \hline
\end{tabular}
\end{table}

\begin{figure*}[!t]
    \centering
    \includegraphics[width=0.99\linewidth]{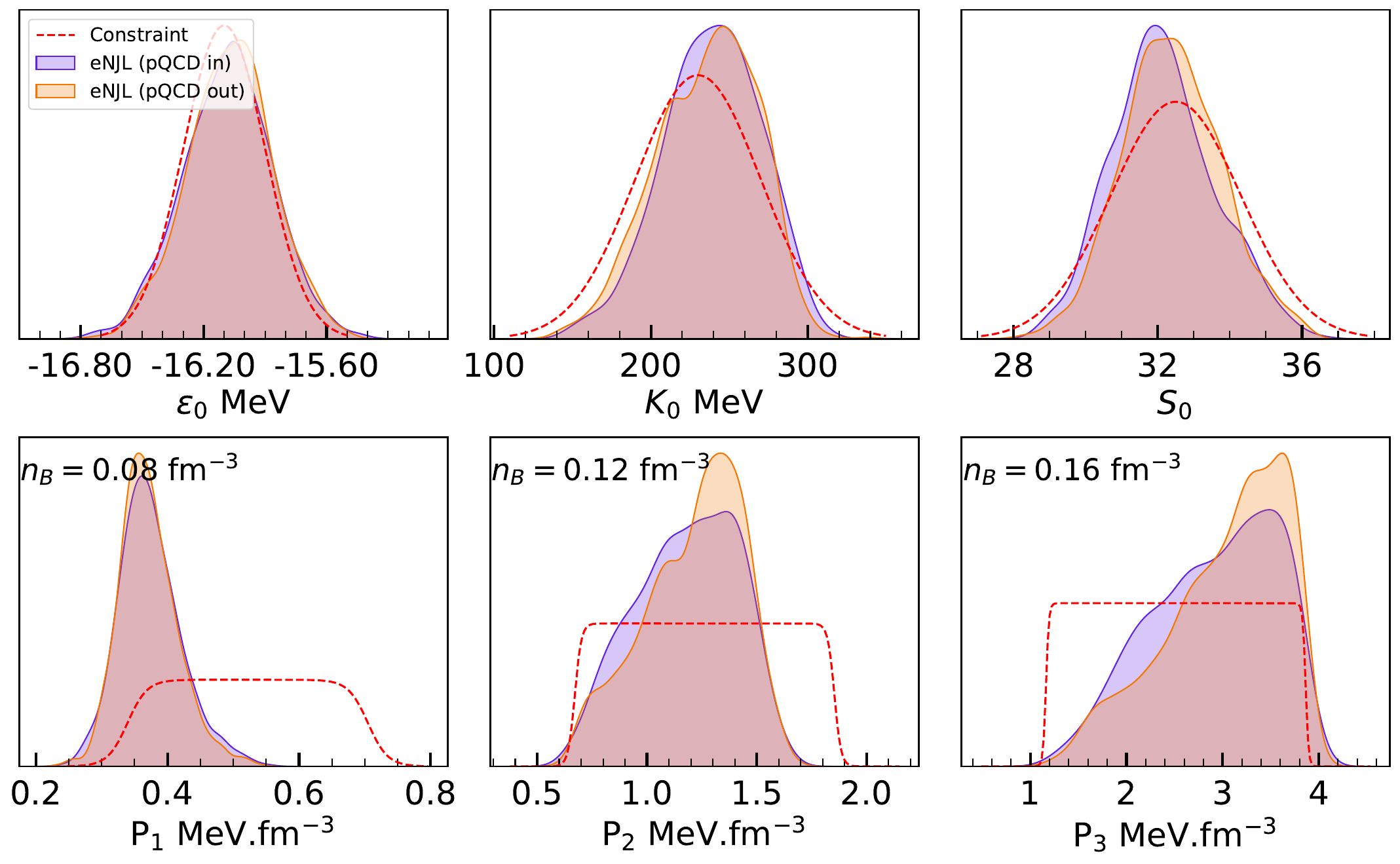}
    \caption{Comparison between the imposed constraints and the associated marginalized posteriors derived from Bayesian inference analysis for models that adhere to the pQCD criteria (pQCD in) or deviate from it (pQCD out).} 
    \label{fig:target}
\end{figure*}

\begin{table*}[!t]
\caption{The nucleon current mass (in MeV) and couplings parameters median values and 90\% CI limits inferred by our Bayesian analysis, displayed for the set with/without the pQCD constraints. }
\label{tab:couplings}
     \setlength{\tabcolsep}{12.pt}
\renewcommand{\arraystretch}{1.4}
\begin{tabular}{ccccccc}
\hline \hline 
Parameter & \multicolumn{3}{c}{pQCD in} & \multicolumn{3}{c}{pQCD out} \\ 
& Median & -90\% CI & +90\% CI & Median & -90\% CI & +90\% CI \\ \hline
$m$ & 427.44 & 308.84 & 480.59 & 358.23 & 185.43 & 452.60 \\ 
$G_s$ & 3.61 & 3.08 & 4.04 & 4.76 & 3.95 & 6.15 \\ 
$G_v$ & 2.78 & 2.49 & 3.13 & 3.84 & 3.11 & 5.04 \\ 
$G_{sv}$ & -9.78 & -13.18 & -5.64 & -12.15 & -17.86 & -7.17 \\ 
$G_\rho$ & 0.23 & -0.11 & 0.43 & 0.32 & 0.06 & 0.43 \\ 
$G_{v\rho}$ & 1.75 & -0.13 & 4.81 & 1.31 & -0.18 & 4.43 \\ 
$G_{s\rho}$ & 6.14 & 2.17 & 12.24 & 7.52 & 2.62 & 15.14 \\ \hline
\end{tabular}
\end{table*}

\begin{figure}
    \centering
    \includegraphics[width=1\linewidth]{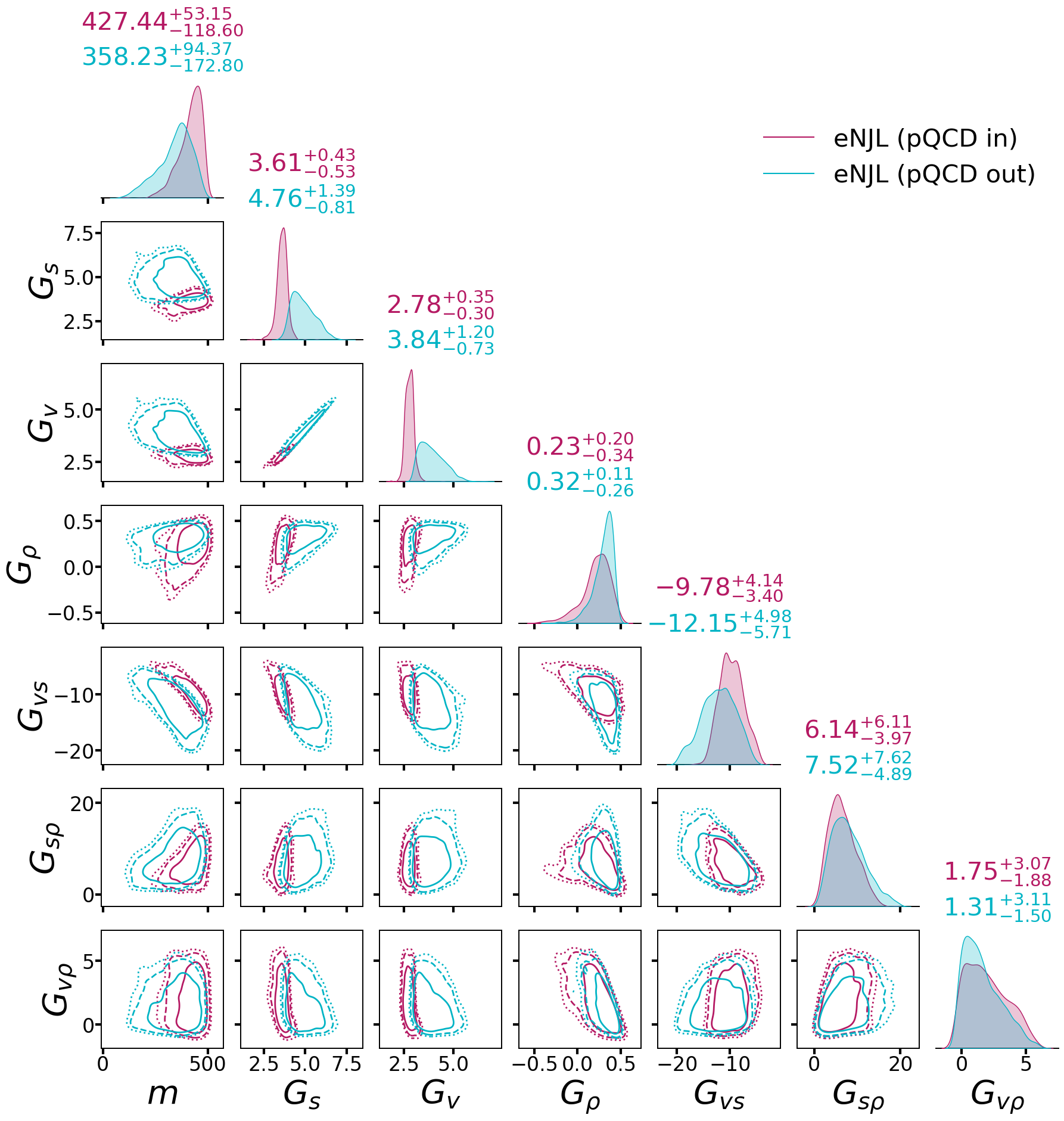}
    \caption{{ A corner plot illustrating the two-dimensional distributions for the eNJL parameters, where the pQCD constraints are applied (in pink, pQCD in) and not applied (in blue, pQCD out). The diagonal plots display the 90\% confidence intervals for the one-dimensional nuclear matter properties. The figures above correspond to the median and the boundaries of the 90\% confidence intervals. The curves indicate the 68\% (full line), 95\% (dashed line) and, 99\% (dotted line) confidence intervals.}}
    \label{fig:parameters}
\end{figure}

\begin{figure*}[!t]
    \centering
    \includegraphics[width=0.48\linewidth]{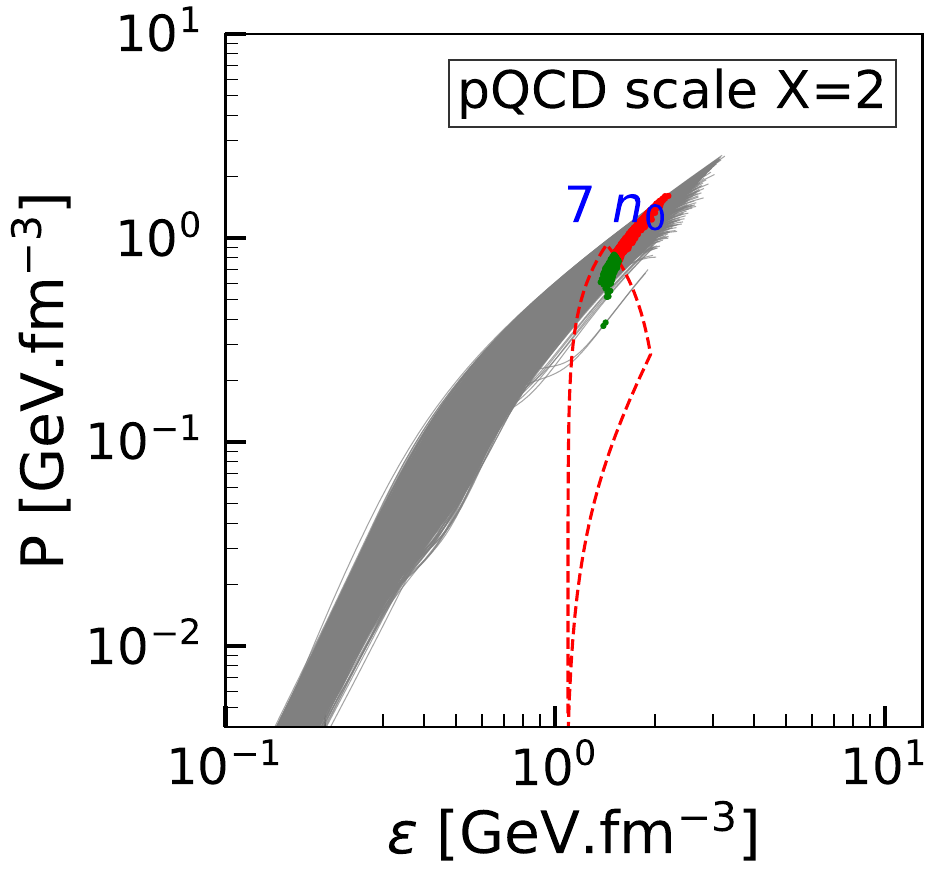}
    \includegraphics[width=0.47\linewidth]{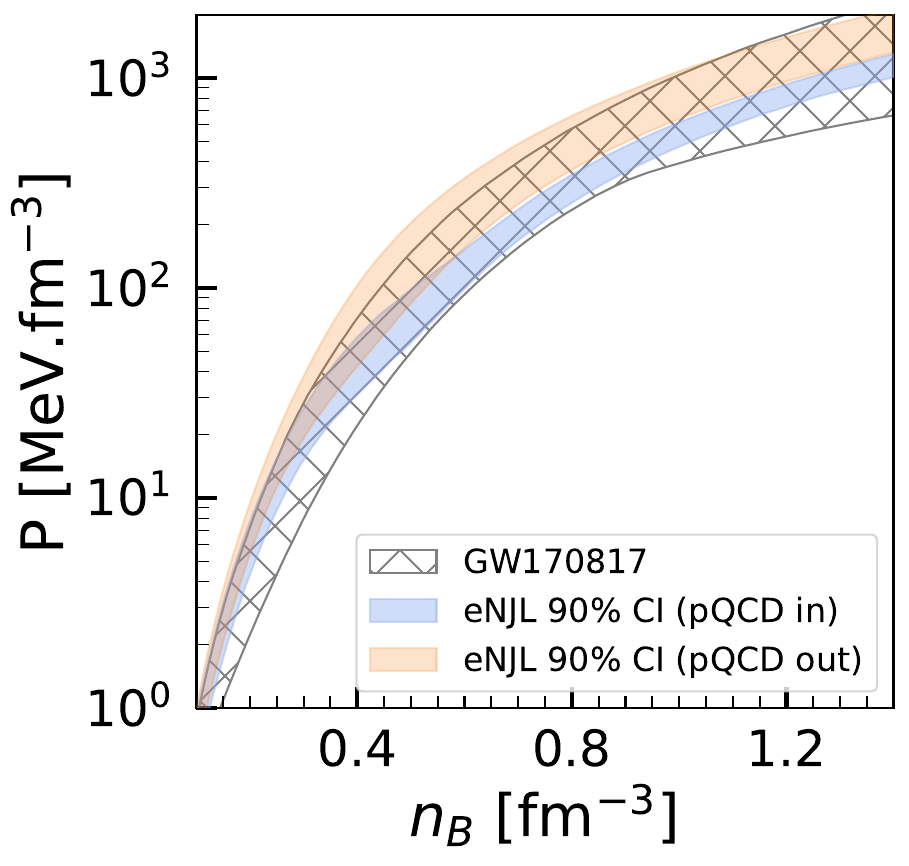}
    \caption{Inferred NS matter (chage neutrality and $\beta$-equilibrium) EoS within the hadronic eNJL model formalism. (left) Pressure versus energy density. The red dashed line delineates the region constrained by considering the pQCD EoS at high densities together with thermodynamic relations and causality \cite{Komoltsev:2021jzg}, for the values of $P$ and $\varepsilon$ at a baryon number density of $n_B = 7n_0$ (where $n_0=0.16$fm$^{-3}$ is the nuclear saturation density). The green and red dots represent the point where $n_B = 7n_0$ for each curve, respectively the models that satisfy or do not satisfy the pQCD constraints. The pQCD constraints were considered for the renormalization scale parameter $X=2$ \cite{Kurkela:2009gj};    (right) The 90\% confidence intervals (CIs) for the pressure of neutron star matter as a function of the baryon density $n_B$, considering models that respect (pQCD in) and violate (pQCD out) the pQCD constraints. Additionally, we include the pressure constraints from GW170817, for comparison \cite{LIGOScientific:2018cki}.}
    \label{fig:eos_pQCD}
\end{figure*}

We calculate the nuclear, pure neutron, and stellar matter properties of the models described by the Lagrangian density presented in Sec. \ref{sec11} by varying its parameters (i.e., the coupling constants and the nucleon bare mass). To infer the parameter fit of the model, we have imposed (i) experimental constraints for symmetric nuclear matter (SNM) at saturation, in particular the binding energy per nucleon, the incompressibility modulus, and the symmetry energy \cite{Dutra2014,MUSES:2023hyz}; (ii) constraints from {\it ab-initio} $\chi$EFT calculations for pure neutron matter (PNM) \cite{Hebeler:2013nza} at three different values of the baryon density; (iii) that the model supports neutron stars with at least $2 ~\rm M_\odot$ to describe the supermassive pulsars PSR J1614-2230 and PSR J0348+0432 \cite{Demorest2010,Antoniadis2013}; (iv) constraints on the hadronic EoS derived from the pQCD EoS at very high density, through the application of thermodynamic relations and the imposition of causality, as explained in Ref. \cite{Komoltsev:2021jzg}; and (v)  $\partial \mathcal{E}/\partial n_B>0$ for $n_B>n_0$ and {${\cal S}(n_B)>0$ in SNM, and $\partial P/\partial n_B>0$ in PNM for all the baryon density range considered.} {These constraints are summarized  in Table \ref{tab:constraints}}.
 
The models generated in our study are categorized according to whether they satisfy the pQCD constraint at a baryon number density of $n_B = 7n_0$ (where $n_0=0.16$ fm$^{-3}$ is the nuclear saturation density) or not. This reference density was chosen because it characterizes the largest densities found within NS. This constraint is shown in the left panel of Fig.~\ref{fig:eos_pQCD} and is incorporated into the likelihood as a filter function, i.e., a given EoS has a probability equal to one if that point is inside this region and zero if it is outside. We denote the (non)inclusion of the constraint as {'pQCD (out)in'}. Notice that the pQCD EoS depends on the renormalization scale $X$, with $X=2$ being its central value \cite{Kurkela:2009gj}, the value considered in the present study when the pQCD constraints are imposed.

 Fig.~\ref{fig:target} shows the experimental constraints imposed in our Bayesian inference, assuming a Gaussian distribution given by Eq. \eqref{eq:gausslikelihood} in the case of the SNM quantities and the super-Gaussian box function given by Eq. \eqref{eq:boxlikelihood} in the case of the $\chi$EFT PNM constraints, compared to the distributions of the nuclear and pure neutron matter quantities obtained for the models generated in our study, for the pQCD in and out categories.
  We find that the experimental constraints on nuclear matter are very well reproduced for both categories of models. For the $\chi$EFT constraints, the models tend to lower pressures, at $n_B\sim n_0/2\sim0.08$ fm$^{-3}$, and the opposite happens at $n_B\sim n_0$, but they are still within the allowed band.  It can be seen that the removal of the 'pQCD out' models shifts the symmetry energy to smaller values and the incompressibility modulus to slightly larger values. Also the PNM pressure marginalized posteriors are a little more spread within the allowed band for the 'pQCD in' category, while especially for the higher $n_B$ points the 'pQCD out' sets are more biased towards larger values. 
  
  The model parameters resulting from the Bayesian inference are given in Table \ref{tab:couplings}. It is interesting to note that the median current mass is above 400 MeV when pQCD constraints are imposed, consistent with the finite current mass parametrizations proposed in Ref. \cite{enjl}, but much larger than the mass of the parametrizations in Ref. \cite{Wei:2015aep}, which proposes sets with all $m$ below 80 MeV. A main effect of including the pQCD constraints is a $\sim 25-30\%$ reduction of the $G_s$, $G_v$ couplings and a smaller variation of all other couplings together with an increase of the current nucleon mass of about $20\%$. { This is clearly seen in the corner plot shown in Fig. \ref{fig:parameters}, where the parameters of the model resulting from the inference calculation are given with and without the pQCD constraints. The main effect of the pQCD constraints is to reduce the magnitude of the coupling $G_v$. Since there is a strong correlation between $G_s$ and $G_v$, the parameter $G_s$ suffers an equally large reduction. The pQCD constraints also noticeably affect the current mass, whose distribution is pushed to the values above 250 MeV.}

In the right panel of Fig.~\ref{fig:eos_pQCD} we plot the pressure against the baryon density for the resulting EoS, with and without the pQCD constraint.  Both sets of models are essentially within the allowable range inferred from gravitational wave and electromagnetic observations of the GW170817 neutron star merger event \cite{LIGOScientific:2018cki,Li:2021thg}, especially in the high-density region. Models that do not satisfy the pQCD constraint are stiffer, as also noted in the left panel, and tend to be slightly off the observational band.

The corner plot in Fig. \ref{fig:corner} summarizes the nuclear matter properties of the EoS dataset resulting from the inference process. We include both results imposing the pQCD constraints and not imposing these constraints. Overall, most of the properties are only slightly affected, with the exceptions being the effective nucleon mass $M^*$, which is way more restricted to higher values in the 'pQCD in' scenario,
and the kurtosis coefficients $Z_0$ and $Z_{\rm sym,0}$. These last two represent the fourth-order parameters in the Taylor expansion of the binding energy and symmetry energy, respectively, and as so, it is expected that a high-density constraint reflects itself on the higher-order terms, hindering large values that would make the EoS very stiff. This is true for both the isoscalar and isovector channels. The other property strongly affected is the effective mass at saturation: pQCD constraints limit $M^*$ to values of the order {(0.75--0.81)$M_0$} 
while, if this condition is removed, values as low as {0.6$M_0$} 
may be possible. This implies that pQCD conditions favor a slow restoration of the chiral symmetry with respect to density. 

\begin{figure*}[!t]
    \centering
    \includegraphics[width=1\linewidth]{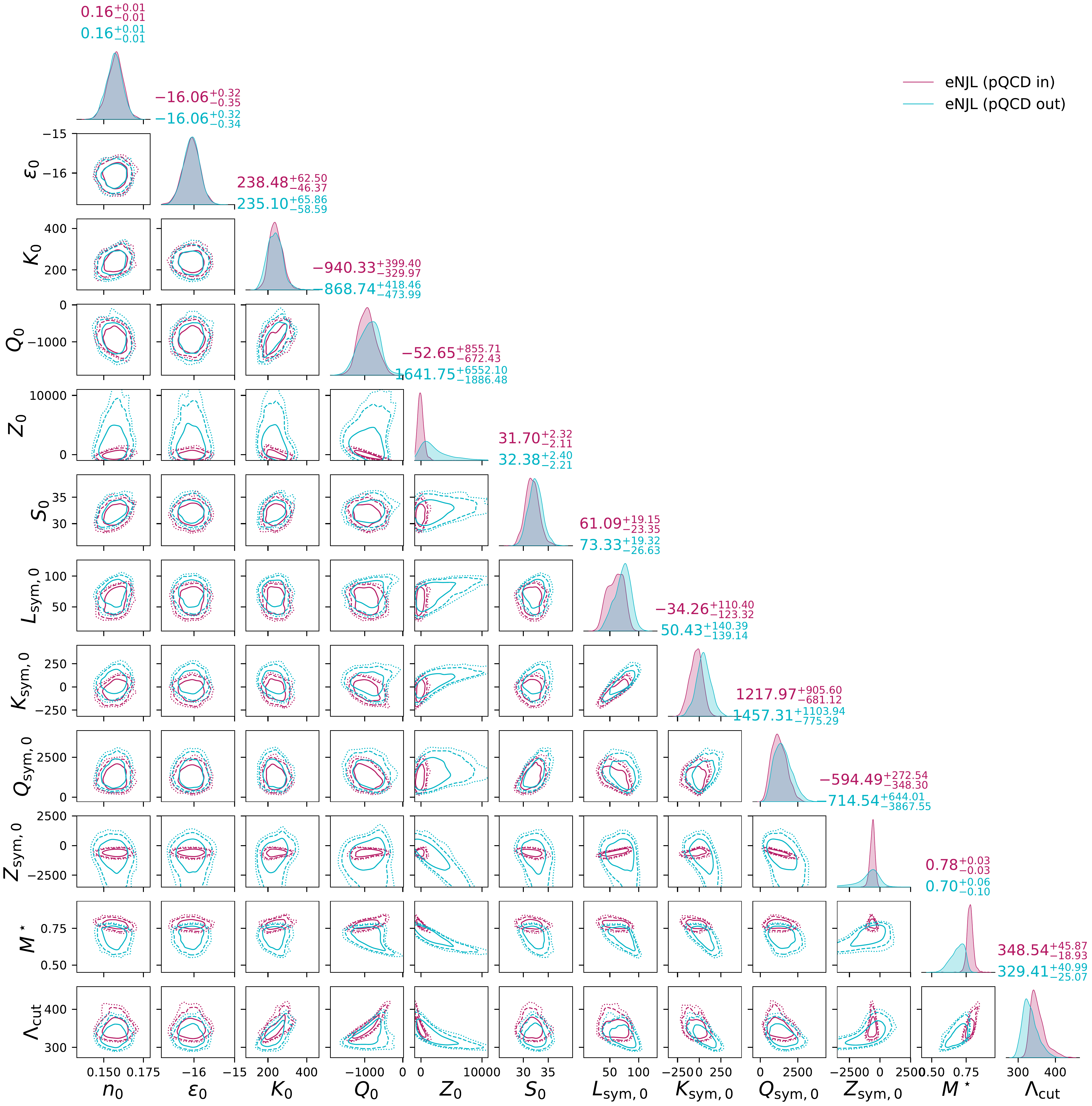}
    \caption{Corner plot showing the 2D distributions for the nuclear matter properties, imposing (pink, pQCD in) and not imposing (blue, pQCD out) the pQCD constraints. The diagonal plots contain the 90\% CI for the 1D nuclear matter properties. The numbers on top refer to the median and 90\% CI limits. The effective nucleon mass {normalized to the bare nucleon mass} $M^*/M_0$ and the cut-off parameter $\Lambda_{\rm cut}$ have also been included. All quantities are in MeV, {except for the effective mass ratio}. { The curves indicate the 68\% (full line), 95\% (dashed line) and, 99\% (dotted line) CI.}}
    \label{fig:corner}
\end{figure*}

\begin{table*}[!ht]
 \caption{List of nuclear matter properties of the dataset obtained: nuclear saturation density $n_0$ (in fm$^{-3}$), binding energy $\varepsilon_0$, incompressibility modulus $K_0$, skewness $Q_0$, kurtosis $Z_0$, symmetry energy $\mathcal{S}_0$, its slope $L_{\rm sym,0}$, curvature $K_{\rm sym,0}$, skewness $Q_{\rm sym,0}$, and kurtosis $Z_{\rm sym,0}$ (all in MeV), the normalized nucleon effective mass $M^*/M_0$, and the cutoff $\Lambda_{\rm cut}$ { (in MeV), that is calculated imposing that the vacuum nucleon mass to be $M_0=939$ MeV}. The resulting compact stars properties are also shown: the maximum star mass $M_{\rm max}$ (in $M_\odot$), {its radius $R_{\rm max}$} (in km) and {its tidal deformability parameter $\Lambda_{\rm max}$}, its central baryon density $n_{B,\rm max}$ (in fm$^{-3}$), energy density $\varepsilon_{\rm max}$ (in MeV.fm$^{-3}$), speed of sound $c^2_{s,\rm max}$, polytropic index $\gamma_{\rm max}$ and the parameter $d_{\rm max}$. The radii of the 1.2, 1.4, 1.8, and $2.0 ~\rm M_\odot$ stars are also given, together with their respective tidal deformabilities. 
 }
 \label{tab:prop}
     \setlength{\tabcolsep}{12.pt}
\renewcommand{\arraystretch}{1.4}
\begin{tabular}{ccccccc}
\hline \hline 
\multirow{2}{*}{Parameter} & \multicolumn{3}{c}{pQCD in}   & \multicolumn{3}{c}{pQCD out}  \\ \cline{2-7}  & Median  & -90\% CI & +90\% CI & Median  & -90\% CI & +90\% CI \\ \hline
$n_0$ & 0.16 & 0.15 & 0.17 & 0.16 & 0.15 & 0.17 \\ 
$\epsilon_0$ & -16.06 & -16.41 & -15.74 & -16.06 & -16.39 & -15.74 \\ 
$K_0$ & 238.48 & 192.11 & 300.98 & 235.10 & 176.50 & 300.95 \\ 
$Q_0$ & -940.33 & -1270.30 & -540.94 & -868.74 & -1342.73 & -450.28 \\ 
$Z_0$ & -52.65 & -725.09 & 803.06 & 1641.75 & -244.73 & 8193.85 \\ 
$\mathcal{S}_0$  & 31.70 & 29.59 & 34.03 & 32.38 & 30.17 & 34.78 \\ 
$L_{\rm sym,0}$ & 61.09 & 37.75 & 80.24 & 73.33 & 46.70 & 92.65 \\ 
$K_{\rm sym,0}$ & -34.26 & -157.58 & 76.13 & 50.43 & -88.72 & 190.82 \\ 
$Q_{\rm sym,0}$ & 1217.97 & 536.85 & 2123.57 & 1457.31 & 682.02 & 2561.25 \\ 
$Z_{\rm sym,0}$ & -594.49 & -942.78 & -321.95 & -714.54 & -4582.09 & -70.53 \\ 
$M^*/m$ & 0.78 & 0.74 & 0.81 & 0.70 & 0.60 & 0.75 \\ 
$\Lambda_{\rm cut}$ & 348.54 & 329.60 & 394.41 & 329.41 & 304.34 & 370.40 \\  \hline
$M_{\rm max}$ & 2.10 & 2.01 & 2.18 & 2.42 & 2.21 & 2.74 \\ 
$R_{\rm max}$ & 10.47 & 9.90 & 11.26 & 11.70 & 10.79 & 12.96 \\ 
$\Lambda_{\rm max}$ & 5 & 4 & 8 & 4 & 3 & 6 \\ 
$n_{\rm B, max}$ & 1.08 & 0.96 & 1.21 & 0.85 & 0.67 & 1.01 \\ 
$\varepsilon_{\rm max}$ & 1413.52 & 1227.54 & 1588.11 & 1091.63 & 852.64 & 1312.69 \\ 
$c_{\rm s, max}^2$ & 0.75 & 0.68 & 0.82 & 0.80 & 0.74 & 0.83 \\
$\gamma_{\rm max}$ & 1.58 & 1.50 & 1.72 & 1.59 & 1.54 & 1.62 \\
$d_{c,\rm max}$ & 0.31 & 0.27 & 0.35 & 0.34 & 0.31 & 0.36 \\ \hline 
$R_{1.4}$ & 12.41 & 11.48 & 13.20 & 13.21 & 12.36 & 13.96 \\ 
$R_{1.6}$ & 12.14 & 11.23 & 13.13 & 13.18 & 12.23 & 14.04 \\ 
$R_{1.8}$ & 11.78 & 10.90 & 12.92 & 13.09 & 12.03 & 14.09 \\ 
$R_{2.0}$ & 11.24 & 10.13 & 12.47 & 12.93 & 11.77 & 14.09 \\ 
$\Lambda_{1.4}$ & 436 & 242 & 706 & 697 & 431 & 1063 \\ 
$\Lambda_{1.6}$ & 157 & 87 & 295 & 301 & 170 & 493 \\ 
$\Lambda_{1.8}$ & 55 & 29 & 118 & 132 & 68 & 240 \\ 
$\Lambda_{2.0}$ & 16 & 6 & 40 & 58 & 25 & 119 \\ 
 \hline
\end{tabular}
\end{table*}

The initial section of Table \ref{tab:prop} presents the calculation of various properties of nuclear matter along with the median value and the confidence level of 90\%. First, we discuss the results imposing the pQCD constraints, to compare our results with the best estimates available in the literature. In our work, the incompressibility modulus takes values between 220 and 300 MeV, which is within the accepted range according to different studies \cite{Khan:2012ps,Stone2014,Huth:2020ozf}. Regarding the symmetry energy and its slope, we obtained $\mathcal{S}_0=31.7^{+ 2.33}_{- 2.11}$ MeV and $L_{\rm sym,0}=61.09^{+ 19.15}_{ - 23.34}$ MeV, in very good agreement with the ranges obtained by \cite{Oertel:2016bki} from a compilation of several experimental and observational constraints, which yielded $\mathcal{S}_0=31.7 \pm 3.2$~MeV and $L_{\rm sym,0}=58.7 \pm 28.1$~MeV, respectively. The symmetry energy curvature $K_{\rm sym,0}$ takes values between $-157.6$ and 76.1 MeV here, allowing positive values that were not obtained in the study of \cite{Li:2021thg}, where $K_{\rm sym,0} = - 107 \pm 88$ MeV. Positive values of $K_{\rm sym,0}$ were proposed in \cite{Reed:2023cap} to describe the CREX and PREX2 results, and were also not excluded in \cite{Zhang:2018vrx}, where both experimental and observational constraints were analyzed and an upper limit of about 68 MeV was proposed.

The NS $M(R)$ and $\Lambda(M)$ curves were obtained by integrating Eqs. (\ref{eq:tov}) and (\ref{eq:tovm}) together with the tidal deformability equations derived in \cite{Hinderer:2007mb}. In Fig.~\ref{fig:mr} we show the 90\% confidence interval for the mass-radius curve $M(R)$ (left) and the tidal deformability $\Lambda(M)$ (right). The {(orange) red dotted region defines the 90\% CI when the pQCD constraints are (not) included}. These two categories of models {partly} overlap:  
those satisfying the pQCD constraint generally predict smaller radii and smaller maximum masses. In particular, the pQCD constraints exclude at 90\% CI all stars with radii $R_{1.4}\gtrsim13.2$ km 
{for NS with mass $1.4$ M$_\odot$} and maximum masses above {2.3} M$_\odot$, see Table \ref{tab:prop} for a summary of the main nuclear matter and neutron star properties. In Fig.~\ref{fig:mr} we have also included the mass-radius band spanned by non-linear RMF models in \cite{Malik:2023mnx}. For low and middle mass stars, the eNJL model is able to cover a larger region of the mass-radius diagram, both on the low and high radius sides. Also, larger maximum masses are obtained within eNJL. The figure also includes some observational data: the NICER constraints for pulsars PSR J0030+0451 \cite{Riley2019,Miller2019}, PSR J0740+6620 \cite{Riley2021,Miller2021}, the recently announced mass and radius of pulsar PSR J0437-4715 \cite{Choudhury24} (median and 1$\sigma$ uncertainty) and the mass-radius contours derived from the GW170817 detection \cite{LIGOScientific:2018cki}.  Note that all models appear to fall within both the NICER and GW170817 contours, except for PSR J0437-4715, whose median lies outside the 90\% contours.

\begin{figure*}
    \centering
\includegraphics[width=1.0\linewidth]{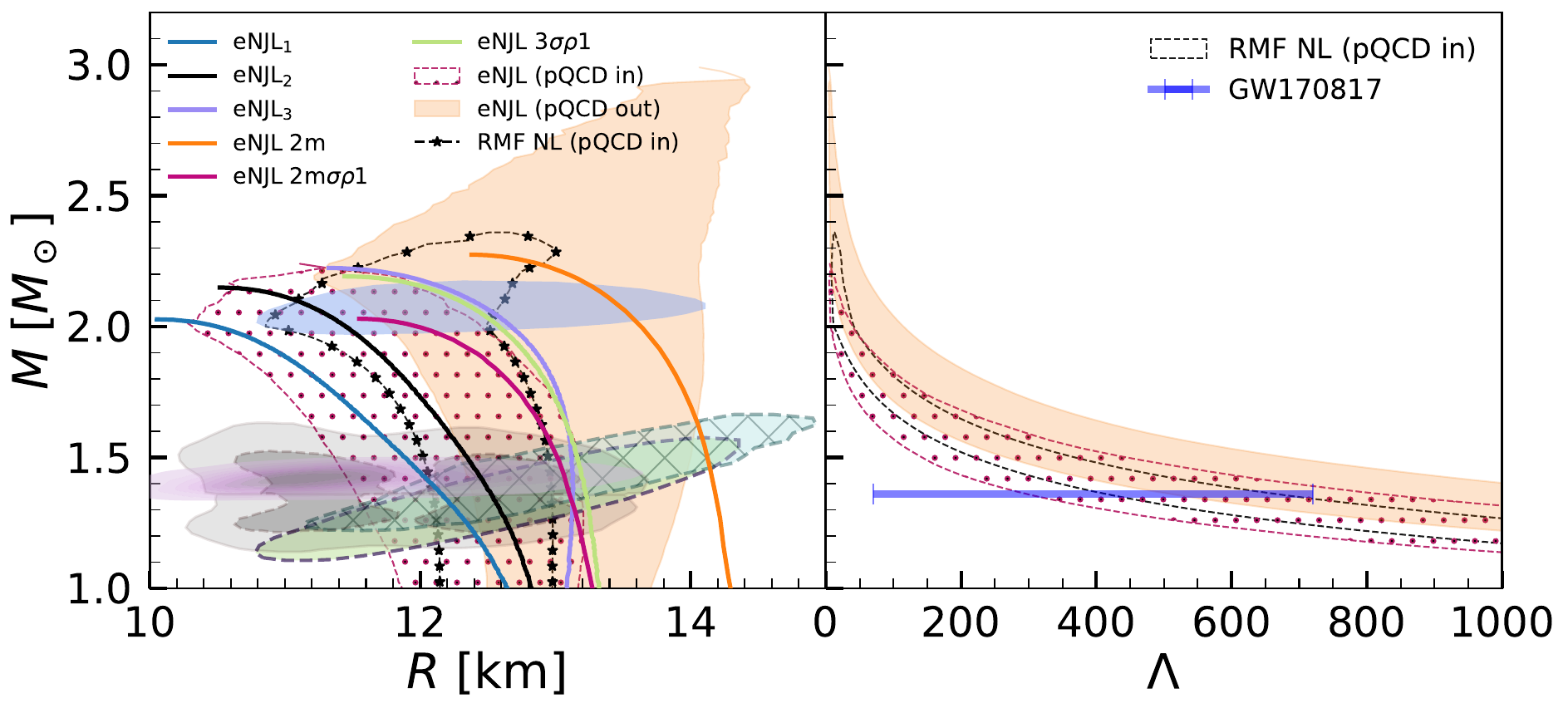}
    \caption{The 90\% confidence region of the mass-radius relations (left) and the mass-dimensionless tidal deformability (right) for models conforming to (pQCD in) and deviating from (pQCD out) pQCD constraints at 7 times the saturation density, with the pQCD renormalizable scale set to $X=2$. {The inferred region obtained by Ref.~\cite{Malik:2023mnx} using a non-linear RMF model framework is included for comparison.} In the left panel, the gray area represents the constraints derived from the binary components of GW170817, along with their corresponding 90\% and 50\% CI \cite{LIGOScientific:2018cki}. Also shown are the 68\% CI for the 2D mass-radius posterior distribution of the millisecond pulsars PSR J0030+0451 (cyan and yellow) \cite{Riley2019,Miller2019} and PSR J0740+6620 (purple) \cite{Riley2021,Miller2021} from the NICER X-ray data. In addition, we show the latest NICER mass and radius measurements for PSR J0437-4715 (silver) \cite{Choudhury24}. The mass-radius curves for the sample EoS eNJL$_{1,2,3}$ were included together with EoS eNJL2m, eNJL2m$\sigma\rho$1 and eNJL3$\sigma\rho$1 obtained from Ref. \cite{enjl}. In the right panel, the blue bar shows the tidal deformability constraints at 1.36 M$_\odot$ \cite{LIGOScientific:2018cki}.
    \\}
    \label{fig:mr}
\end{figure*} 

\begin{table}[!t]
\caption{Parameters of three sample posterior EoS models- $\Lambda_{\rm cut}$ is a derived quantity imposing the nucleon vacuum mass to be 939 MeV.}
\label{tab:eNJLi1}
     \setlength{\tabcolsep}{12.pt}
\renewcommand{\arraystretch}{1.4}
\begin{tabular}{cccc}
\hline \hline 
Parameter & eNJL$_1$ & eNJL$_2$ & eNJL$_3$ \\ \hline
$m$	&	459.70	&	430.97	&	290.80  \\ 
$G_s$	&	3.386	&	3.766	&	3.492 \\ 
$G_v$	&	2.532	&	2.883	&	2.955 \\ 
$G_{sv}$	&	-9.515	&	-10.806	&	-6.357 \\ 
$G_\rho$	&	0.263	&	0.257	&	-0.026 \\ 
$G_{v \rho}$	&	1.255	&	1.084	&	2.906 \\ 
$G_{s \rho}$	&	8.426	&	6.881	&	4.994 \\ 
$\Lambda_{\rm cut}$	&	348.78	&	343.10	&	382.73 \\ \hline 
\end{tabular} 
\end{table}

{We also include in the left panel of Fig. \ref{fig:mr} the $M(R)$ curves of three inferred EoS sample models that yield a canonical star radius $R_{1.4}$ close to the median and the extremes at 90\% CI of the posterior set satisfying pQCD constraints, labeled eNJL$_1$, eNJL$_2$, and eNJL$_3$. Their parameters are given in Table \ref{tab:eNJLi1} and the nuclear matter and neutron star properties in Table \ref{tab:eNJLi2}. For these parameterizations, a unified inner crust-core EoS were built according to the approach discussed in \cite{enjl}. These models complete EoS will be uploaded in the {zenodo database \cite{eNJL_uni}}.
Some noteworthy properties that distinguish them are (i) the fact that eNJL$_3$ has rather larger incompressibility $K_0$ and skewness $Q_0$ coefficients, but a smaller kurtosis $Z_0$ in comparison to the other sets; (ii) regarding the symmetry energy properties, eNJL$_1$ has larger symmetry energy at saturation but a smaller curvature $K_{\rm sym,0}$; and (iii) noticing the overall behavior of the radii and tidal deformability is an increasing trend from eNJL$_1$ to eNJL$_3$. For comparison, we include {three} models taken from Ref.~\cite{enjl}, labeled as eNJL3$\sigma\rho1$, {eNJL2m} and eNJL2m$\sigma\rho1$. These models can describe 2 M$_\odot$ stars, {but eNJL2m does not fulfill the $\chi$EFT constraints from \cite{Hebeler:2013nza}, and the other two can}  only partly fulfill them, deviating for densities above $n_B\gtrsim 0.15$ fm$^{-3}$. They predict larger radii for low- and intermediate-mass stars, and are barely within the 'pQCD in' band obtained here, {except for eNJL2m that is completely outside that band, as it has extremely large radii}. These improved results serve to illustrate the usefulness of the Bayesian inference methodology to the fitting of the theoretical models according to experimental and observational constraints.}

\begin{figure}[!t]
    \centering
    \includegraphics[width=1\linewidth]{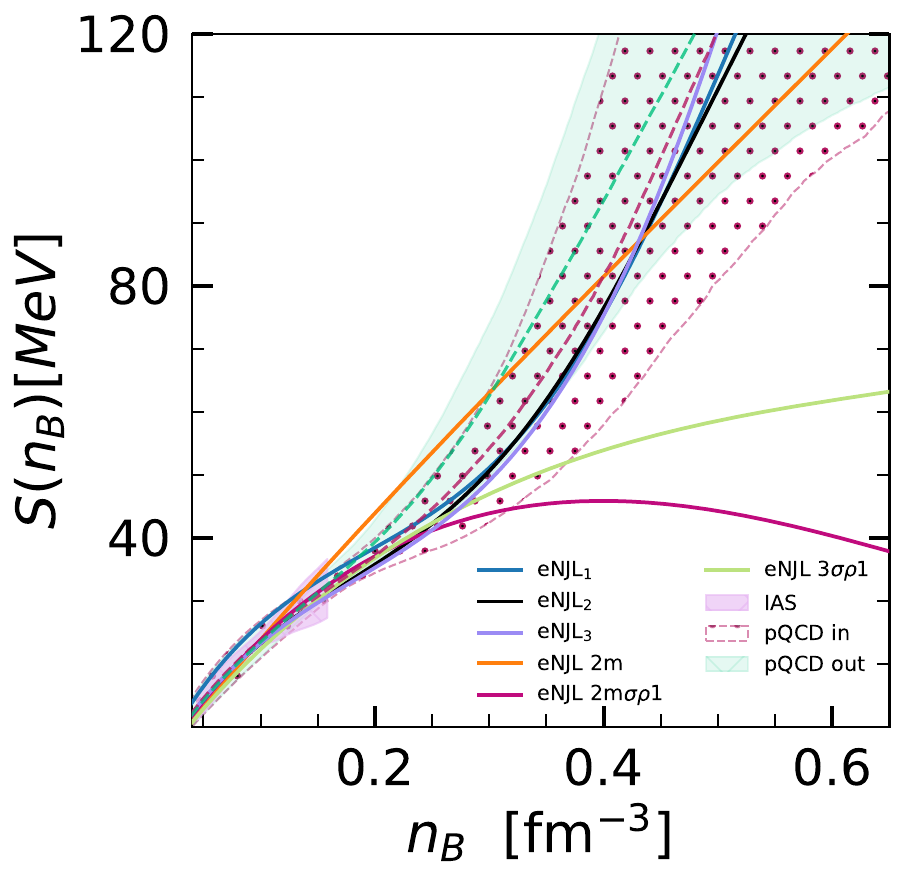}
    \caption{The nuclear matte symmetry energy $\mathcal{S}$ as function of the baryon density $Sn_B$. The red dotted (green) band represents the inferred eNJL models inside (outside) pQCD constraints. The purple region represents the IAS heavy-ion collision flow data on the pressure of symmetric nuclear matter. The symmetry energy curves for the sample eNJL$_{1,2,3}$ models are also included, together with three models obtained from Ref. \cite{enjl} (eNJL2m, eNJL2m$\sigma\rho$1 and eNJL3$\sigma\rho1$), for comparison.}
    \label{fig:esym}
\end{figure}

\begin{table}
 \caption{Nuclear matter and neutron star properties of three sample posterior EoS.}
 \label{tab:eNJLi2}
     \setlength{\tabcolsep}{12.pt}
\renewcommand{\arraystretch}{1.4}
\begin{tabular}{cccc}
\hline \hline 
{Parameter} & eNJL$_1$ & eNJL$_2$ & eNJL$_3$ \\ \hline
$n_0$	&	0.161	&	0.155	&	0.158	 \\ 
$\mathcal{E}_0$	&	-15.71	&	-16.05	&	-15.86	 \\ 
$K_0$	&	232	&	237	&	285	 \\ 
$Q_0$	&	-998	&	-999	&	-621	 \\ 
$Z_0$	&	-85	&	69	&	-572	 \\ 
$\mathcal{S}_0$	&	34	&	31	&	31	 \\ 
$L_{\rm sym,0}$	&	53	&	55	&	53	 \\ 
$K_{\rm sym,0}$	&	-99	&	-55	&	-63	 \\ 
$Q_{\rm sym,0}$	&	1472	&	1145	&	1205	 \\ 
$Z_{\rm sym,0}$	&	-782	&	-565	&	-527	 \\ 
$M^*/m$	&	0.788	&	0.771	&	0.778	 \\ 
\hline 
$M_{\rm max}$	&	2.019	&	2.140	&	2.215	\\
$R_{\rm max}$	&	10.01	&	10.44	&	11.26	\\
$\Lambda_{\rm max}$	&	5	&	5	&	7	\\
$n_{\rm B,max}$	&	1.191	&	1.077	&	0.941	\\
$\varepsilon_{\rm max}$	&	1560	&	1407	&	1197	\\
$c_{\rm s, max}^2$	&	0.763	&	0.783	&	0.717	\\
$\gamma_{max}$	&	1.548	&	1.556	&	1.659	\\
$d_{\rm c, max}$	&	0.314	&	0.327	&	0.301	\\ \hline
$R_{1.4}$	&	11.89	&	12.20	&	12.83	\\
$R_{1.6}$	&	11.52	&	11.96	&	12.83	\\
$R_{1.8}$	&	11.09	&	11.68	&	12.74	\\
$R_{2.0}$	&	10.32	&	11.26	&	12.46	\\
$\Lambda_{1.4}$	&	288	&	391	&	596	\\
$\Lambda_{1.6}$	&	96	&	144	&	259	\\
$\Lambda_{1.8}$	&	32	&	52	&	111	\\
$\Lambda_{2.0}$	&	7	&	17	&	43	\\
 \hline
\end{tabular}
\end{table}

\begin{figure*}
    \centering
    \includegraphics[width=1\linewidth]{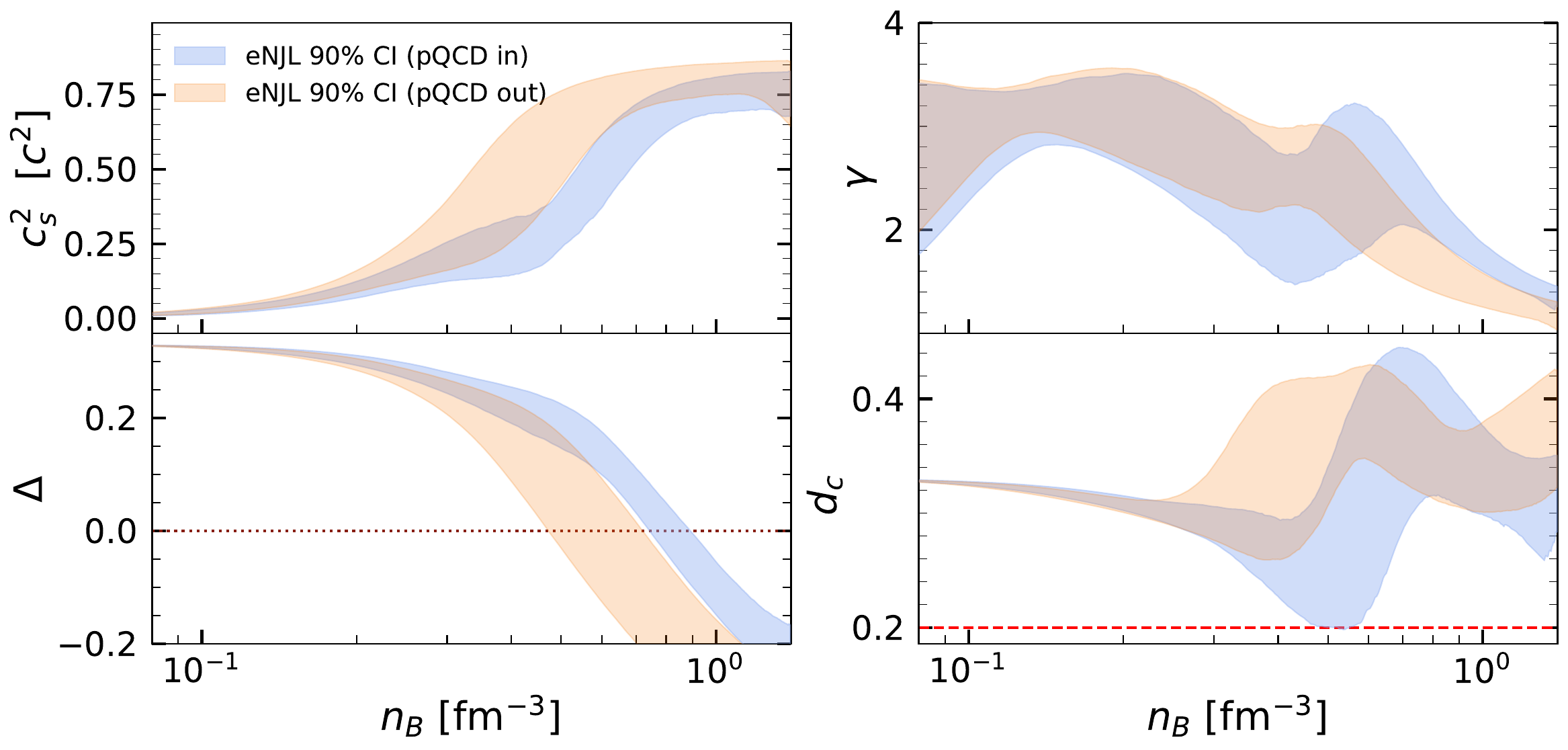}
    \caption{Inferred eNJL models 90\% confidence interval values for (top left) the square of the speed of sound $c_s^2$; (top right) the polytropic index $\gamma=\partial \ln{P}/\partial \ln{\varepsilon}$; (bottom left) the trace anomaly $\Delta=1/3-P/\varepsilon$; and (bottom right) the parameter $d_c=\sqrt{\Delta^2+ (\Delta')^2}$, all as function of baryon density $n_B$ and for both sets of posterior distributions ('pQCD in' and 'pQCD out').}
    \label{fig:cs2_trace}
\end{figure*}

On the right panel of Fig.~\ref{fig:mr} we show the 90\% CI for the tidal deformability as a function of stellar mass. The constraint from GW170817 for the tidal deformability of a 1.36 M$_\odot$ star, corresponding to the merger of two NS of equal mass, is given by a blue horizontal band.
The same trend is observed as in the panel with the $M(R)$ curves, the EoS satisfying the pQCD results tend to have smaller tidal deformabilities, i.e. they are more compact objects, with the 'pQCD in' band slightly overlapping with the band excluded by pQCD. It is also interesting to check that the pQCD constraints exclude most EoS that do not satisfy the GW170817 constraints. Similar to the results covered by the $M(R)$ curves, the non-linear RMF EoS generated in \cite{Malik:2023mnx} also span a narrower band in the case of tidal deformability than the eNJL model proposed here, which is more flexible and able to cover a larger region of the mass-radius diagram. Note that both models, the RMF used in \cite{Malik:2023mnx} and the present eNJL model, contain the same number of parameters to be fitted in the Bayesian inference calculation. 

{In Fig.~\ref{fig:esym}, we show the density dependence of the symmetry energy of the inferred eNJL models, together with three sample models eNJL$_{1,2,3}$ and the parameterizations eNJL2m, eNJL2m$\sigma\rho$1 and eNJL3$\sigma\rho1$, taken from Ref.~\cite{enjl}. The symmetry energy plays a pivotal role in determining the particle distribution as a function of density, especially the fraction of protons. 
For comparison, we present the constraint obtained from heavy-ion collision flow data on the pressure of symmetric nuclear matter \cite{danielewicz02}, which is not entirely model-independent. 
The symmetry energy of the eNJL2m$\sigma\rho1$ and eNJL${3\sigma\rho1}$ models lies outside the posterior distribution we obtained. This discrepancy arises because the former is inconsistent with nuclear matter incompressibility, and the latter conflicts with low-density PNM constraints at a density of 0.16 fm$^{-3}$, both quantities we have used in this study as constraints in the parameters inference. The other model, eNJL2m, lies within the our 'pQCD out' posterior band. This model presents a high slope of the symmetry energy at saturation, $L=89$ MeV. The model eNJL2m$\sigma\rho1$ belongs to this family, i.e., has the same isoscalar properties, but a term that mixes the isoscalar-scalar $\sigma$ and the isovector-vector $\rho$ channels was added in order to lower the symmetry energy. As depicted in the Figure, this $\sigma\rho$ term leads to a reduction in the symmetry energy, and this model has $L=59$ MeV. }

In order to understand the behavior of the model at high densities we have plotted in Fig.~\ref{fig:cs2_trace} several properties frequently discussed in the literature: the speed of sound squared ${c_s}^2=\partial P/\partial\varepsilon$, the polytropic index $\gamma=\partial \ln{P}/\partial \ln{\varepsilon}$, the renormalized matter trace anomaly $\Delta=1/3-p/\varepsilon$ as a function of density, and the measure of conformality proposed in \cite{Annala:2023cwx}, derived from the renormalized trace anomaly introduced in \cite{Fujimoto:2022ohj} and its logarithmic derivative with respect to the energy density, $d_c=\sqrt{\Delta^2+ ({\Delta'})^2}$, with  $\Delta'=c_s^2\left(\frac{1}{\gamma}-1\right)$. 

These quantities have often been discussed as possible parameters identifying the presence of a phase transition to deconfined quark matter \cite{Annala2020,Fujimoto:2022ohj,Annala:2023cwx}. In the high density limit, the conformal limit is expected to be reached and the square of the speed of sound to approach 1/3. The speed of sound squared increases monotonically for both 'pQCD in' and 'pQCD out'. Although the behavior of the set satisfying pQCD is smoother, both sets reach values of the order of 0.8. In the range of densities shown, the renormalized trace anomaly decreases continuously and crosses the zero axis for densities of the order of {$\sim 3n_0$ (pQCD out) or $\sim 5n_0$ (pQCD in)}, showing no trend that could indicate a change in behaviour. Note, however, that the present model cannot be extended to densities corresponding to Fermi momenta larger than the cutoff $\Lambda_{\rm cut}$. For the polytropic index, values below {3.4} are possible. If  pQCD constraints are included,  the $d_c$ parameter decreases until reaching the value 0.2 at about 4$n_0$,  but at larger densities it increases until values above 0.4. It was argued in Ref. \cite{Annala2020} that models that fall below $d_c=0.2$ would have a transition to deconfined quark matter. {Although the studies \cite{Providencia:2023rxc,Malik:2024qjw} showed that this statement should be taken with care, as the authors worked with purely nucleonic models and found $d_c<0.2$ for some of them, here we only get values above this threshold, suggesting that within the eNJL framework purely hadronic NS are characterized by $d_c>0.2$.}
The value of this parameter  is defined by both the trace anomaly and its derivative and, at high densities,  the renormalised trace anomaly crosses the zero axis with a large slope.

\section{Conclusions}\label{conclusion}
A chirally symmetric model has been considered to describe nuclear matter and neutron stars. The model is a generalization of the NJL model proposed in the 1960s by Nambu and Jona-Lasinio \cite{Nambu:1961tp} for nucleons, which includes not only four-point interactions but also eight-point interactions as proposed in \cite{Koch1987}. The eight-point interactions are essential to correctly describe the properties of nuclear matter \cite{enjl,Wei:2015aep}. Within a Bayesian inference calculation, the seven parameters of the model were constrained by some nuclear matter properties at saturation, neutron matter EoS calculated within a $\chi$EFT formalism at saturation and sub-saturation densities, a maximum NS mass of at least 2 M$_\odot$, an increasing pressure with density, a non-negative symmetry energy and the pQCD EoS. 

The model has a current mass term that explicitly breaks the chiral symmetry. The constrained EoS favor quite large values of the current mass, which came out of the order of 400 MeV as in the parameterizations found in \cite{enjl} which include a non-zero current mass, but much larger than the current mass of the parameterizations proposed in \cite{Wei:2015aep}.

Nuclear matter properties such as saturation density, binding energy at saturation, incompressibility modulus and symmetry energy at saturation take expected values. However, it is interesting to note that our EoS favor a symmetry energy slope $L_{\rm sym,0}$ of the order of $61\pm 20$ MeV and a curvature $K_{\rm sym,0}$ that {not only takes negative values, as predicted in the recent review \cite{Li:2021thg} where $K_{\rm sym}=-107 \pm 88$ MeV at a 68\% confidence level, but can also take positive} values in the range $-158\lesssim K_{\rm sym,0}\lesssim 76$ MeV at 90\% CI. These results include pQCD constraints, otherwise $K_{\rm sym,0}$ would preferentially take positive values. While the slope of the symmetry energy agrees with the values proposed in \cite{Li:2021thg}, which were obtained by constraining the symmetry energy using NS observations, the curvature of the symmetry energy takes much larger values than those proposed in the same study. Positive values for the symmetry energy curvature have been obtained in \cite{Reed:2023cap} to describe the PREX2 \cite{Reed2021} and CREX \cite{CREX:2022kgg} results.  The pQCD constraints mainly concern the fourth-order parameters $Z_0$ and $Z_{\rm sym,0}$ and the nucleon effective mass, which takes values in the range $0.7< M^*/M\lesssim 0.8$. Large values of the effective mass are associated with softer EoS at high densities and smaller NS radii \cite{Hornick2018}.   Although the model can not be extended to describe densities as large as the ones covered by the pQCD EoS, it is possible to analyze its effects at densities as the ones existing inside NS considering causality and thermodynamic relations as proposed in \cite{Komoltsev:2021jzg}.

We have studied the NS properties described by the present model. The pQCD EoS imposes strong constraints and excludes very massive stars and large radii. We have verified that the obtained mass-radius distribution covers a larger mass-radius range than that obtained from an RMF with non-linear mesonic terms formalism in \cite{Malik:2023mnx}. In particular, smaller radii are allowed: at 90\% CI, the radius of a 1.4 M$_\odot$ takes on the values $11.95 \lesssim R_{1.4}\lesssim 13.2$ km.  
The eNJL seems to fit better the recently published data from NICER for the pulsar PSR J0437-4715 with a radius of $R=11.36_{-0.63}^{+0.95}$ km and a mass of $M=1.418\pm0.037 ~\rm M_\odot$.

Maximum masses of the order of 2.4 M$_\odot$ are still allowed by pQCD constraints. The tidal deformability obtained for the complete dataset of EoS is compatible with the constraints resulting from GW170817.  The speed of sound squared shows a monotonically increasing trend taking values of the order of $0.7$--$0.8$ in the core of NS. 
{ A similar study to the one presented here was performed in Refs. \cite{han2023plausible,fan2024maximum} using largely the same $\chi$EFT and pQCD constraints, but finding different behaviors for the stellar maximum mass and  the speed of sound in the high-density region. In subsequent studies, it might be interesting to better understand the underlying reason of these differences (i.e., the dependence on the physical model).}

{Three representative EoS have been selected and the unified inner-crust core $\beta$-equilibrium built and will be made available in the CompOSE (CompStar Online Supernovae Equations of State), the  online repository of equations of state \cite{Typel:2013rza,CompOSECoreTeam:2022ddl}, as well as in the zenodo databse \cite{eNJL_uni}.}

\section*{Acknowledgments}
K.D.M. received support from the Conselho Nacional de Desenvolvimento Científico e Tecnológico (CNPq/Brazil) through grant 150751/2022-2 and D.P.M. under grant 303490/2021-7. K.D.M. was also partially supported by Fundação de Amparo à Pesquisa do Estado de São Paulo (FAPESP) under grant 2024/01623-6. 
This work was partially supported by national funds from FCT (Fundação para a Ciência e a Tecnologia, I.P, Portugal) under projects 
UIDB/04564/2020 and UIDP/04564/2020, with DOI identifiers 10.54499/UIDB/04564/2020 and 10.54499/UIDP/04564/2020, respectively, and the project 2022.06460.PTDC with the associated DOI identifier 10.54499/2022.06460.PTDC. H.P. acknowledges the grant 2022.03966.CEECIND (FCT, Portugal) with DOI identifier 10.54499/2022.03966.CEECIND/CP1714/CT0004. This work was produced with the support of Deucalion HPC, Portugal and it was funded by FCT I.P., Portugal, under the project Advanced Computing Project 2024.14108.CPCA.A3,  RNCA (Rede Nacional de Computação Avançada).
It is also part of the project INCT-FNA proc. No. 464898/2014-5.  

\bibliographystyle{apsrev4-1}
\bibliography{bibliography}

\end{document}